\documentstyle[preprint, aps]{revtex}
\newcommand{\beq}[1]{\begin{equation}\label{#1}}
\newcommand{\eeq}{\end{equation}}

\newcommand{\arcosh}{ {\rm arcosh \ }}

\renewcommand{\Im }{ {\rm Im}}
\renewcommand{\Re }{ {\rm Re}}
\begin{document}
\draft
\title{Semiclassical Black Hole States and Entropy }
\author{Thorsten Brotz and Claus Kiefer}
\address{Fakult\"at f\"ur Physik, Universit\"at Freiburg,
Hermann-Herder-Str.3,\\ D-79104 Freiburg, Germany}

\date{\today}
\maketitle

\begin{abstract}
We discuss semiclassical states in quantum gravity corresponding to 
Schwarzschild as well as Reissner-Nordstr\"om black holes. 
We show that reduced quantisation of these models is equivalent to 
Wheeler-DeWitt quantisation with a particular factor ordering. 
We then demonstrate how the entropy of black holes can be 
consistently calculated from these states. While this leads to the
Bekenstein-Hawking entropy in the Schwarzschild and non-extreme 
Reissner-Nordstr\"om cases, the entropy for the extreme 
Reissner-Nord\-str\"om case turns out to be zero.
\end{abstract}
\pacs{ }

The issues of black hole entropy and Hawking radiation play a key 
role in any attempt to quantise the gravitational field. 
For a deeper understanding it is of central importance to provide 
a satisfactory
interpretation of black hole entropy from statistical mechanics.
Recently, progress on this question has been achieved in the context
of string theory \cite{0a}, but the more conservative framework
of quantum general relativity provides interesting insight into this
question, too. In $2+1$ dimensions a statistical interpretation
has been suggested using the Chern-Simons form \cite{0b},
but in $3+1$ dimensions this remains still elusive.

On the level of the semiclassical approximation, black hole entropy
has been discussed in the framework of path integrals \cite{1,2}.
Such a treatment exploits the formal analogy of euclidean path 
integrals in standard quantum field theory to partition sums in 
statistical mechanics. 
The entropy is then calculated as the logarithm of the density 
of states in the partition function which is found from an appropriate
saddle point approximation to the path integral.
To ensure thermodynamical stability, the black hole has to be enclosed
in a spatially finite box (or, alternatively, has to be embedded in an
anti-de Sitter spacetime \cite{7}). If appropriate boundary conditions
are imposed at the wall of the box and at the black hole horizon, the 
partition function can be evaluated, and the black hole
entropy is found {}from the boundary term at the horizon to take the 
Bekenstein-Hawking value $A/4\hbar G$, where $A$ is the area of the 
horizon \cite{2}.

The purpose of the present paper is to investigate how the black hole 
entropy can be consistently found from semiclassical solutions to 
the Wheeler-DeWitt equation in quantum gravity. Since from a 
physical point of view there is a close connection between the WKB 
approximation for wave functionals and the saddle point approximation
for path integrals, this should be possible to achieve. 
Some interesting new aspects will turn out in this discussion which
thus complements the standard treatment
in the path integral context.

In the following we shall consider spherically symmetric
gravitational systems which include the important cases of the 
Schwarzschild and the Reissner-Nordstr\"om black holes. 
A WKB solution of the Wheeler-DeWitt equation for the former case
was given in \cite{3} (see also \cite{4}). On the other hand, a
reduced quantisation was performed in \cite{5} with the mass of 
the black hole as the only remaining configuration variable
(see \cite{6} for an analogous discussion 
in the framework of connection dynamics). 
We shall show the equivalence of reduced quantisation to 
Wheeler-DeWitt quantisation in a particular factor ordering.
We extend the discussion to include the Reissner-Nordstr\"om case,
where we 
present a careful investigation into the notions of `classically 
allowed' and `classically forbidden' regions. Our main point then will be 
the recovery of the black hole entropy {}from the Hamilton-Jacobi
functional in a consistent way. In the course of this discussion it
will turn out in a natural way that the entropy of the extreme
Reissner-Nordstr\"om black hole vanishes.

We start {}from the ADM form of the general spherically symmetric
spacetime metric on the manifold $ I\!\! R \times I\!\! R \times
S^2 $:
\beq{1}
ds^2 = -N^2 dt^2 + \Lambda^2(r,t) (dr + N^r dt)^2 + R^2(r,t) 
d\Omega^2,
\eeq
where $d\Omega^2$ denotes the standard metric on $S^2$, and $N$ 
($N^r$) is the lapse function (shift function).
Inserting the ansatz (1) into the Einstein-Hilbert action and varying
with respect to $N$ and $N^r$ leads to the Hamiltonian constraint and
the radial momentum constraint \cite{3,4,5},
\beq{2}
{\cal H}_G \equiv \frac{G}{2} \frac{\Lambda P_\Lambda^2}{R^2} - G
             \frac{P_\Lambda P_R}{R} + \frac{V_G}{G} \approx 0,
\eeq
\beq{3}
{\cal H}_r \equiv P_R R' - \Lambda P'_\Lambda \approx 0,
\eeq
where the gravitational potential term $V_G$ reads explicitly
\beq{4}
V_G \equiv \frac{R R''}{\Lambda} - \frac{R R' \Lambda'}{\Lambda^2}
           + \frac{R'^2}{2\Lambda} - \frac{\Lambda}{2}.
\eeq
The inclusion of the cosmological constant $\lambda$ is 
straightforward and would lead to an additional term $\lambda
\Lambda R^2/2$ in (\ref{4}). In the following we shall include in
addition a spherically symmetric electromagnetic field \cite{7}.
The corresponding vector potential is written in the form
\beq{5}
A= \phi(r,t) dt + \Gamma(r,t) dr.
\eeq
This, then, leads to the addition of the kinetic term
\beq{6}
{\cal H}_E \equiv \frac{\Lambda P^2_\Gamma}{2R^2}
\eeq
to (\ref{2}), while (\ref{3}) remains unchanged. Furthermore,
variation with respect to the Lagrange multiplier $\phi$ in the 
action leads to the Gauss constraint
\beq{7}
{\cal G} \equiv P'_\Gamma \approx 0.
\eeq
Boundary conditions for all fields are assumed to hold such that all
integrals are well-defined and such that the classical spacetime 
metric is nondegenerate \cite{7}.
Quantisation is then performed in the standard formal manner by 
replacing all momenta with $\hbar/i$ times functional derivatives and
implementing all constraints by acting on wave functionals
$\Psi[\Lambda(r),R(r),\Gamma(r)]$. At this point one normally has to
rely on a particular factor ordering, but for the following results
we do not need to fix this ambiguity.
The electromagnetic part is trivially to solve: Eq.\ (\ref{7})
becomes
\beq{8}
\frac{d}{dr} \frac{\delta \Psi}{\delta \Gamma(r)} = 0,
\eeq 
which is solved by $\Psi = f(\int_{-\infty}^\infty \Gamma(r) dr ) ^.
\psi[\Lambda(r),R(r)]$, where $f$ is an arbitrary differentiable 
function.
Note that the structure of (8) ensures that 
$\delta\Psi /\delta \Gamma(r)$  does not depend explicitly on $r$ and
therefore guarantees that the second derivatives 
$\delta^2\Psi /\delta \Gamma^2(r)$ are well defined. In fact, one 
immediately finds {}from $(\hat{\cal H}_G+\hat{\cal H}_E) \Psi = 0 $
the solutions
\beq{9}
\Psi = e^{\frac{iq}{\hbar} \int_{-\infty}^{\infty} \Gamma dr}
       \psi[\Lambda(r),R(r)],
\eeq
where $\psi$ satisfies the Wheeler-DeWitt equation, which reads with 
`naive' factor ordering,
\beq{10}
\left\{ - \frac{G\hbar^2\Lambda}{2R^2} \frac{\delta^2}{\delta 
\Lambda^2}
+ \frac{G \hbar^2}{R} \frac{\delta^2 }{\delta \Lambda \delta R}
+ \frac{V_G}{G} + \frac{\Lambda q^2}{2 R^2} \right\} \psi = 0.
\eeq
General solutions are found by performing superpositions of the 
states (\ref{9}) with respect to $q$.

The form (\ref{9}) is of course well known from the discussion of 
two-dimensional QED in the functional Schr\"odinger picture \cite{8}.
The role of the `charge' $q$ is there played by the background value 
of the electric field. In analogy to \cite{8}, one might also wish in 
our case to 
study the transformation of $\Psi$ with respect to large gauge 
transformations, $\Psi \longrightarrow \Psi e^{-2\pi i q n}$, with a 
$\theta$-parameter $\theta \equiv 2\pi q$, but we shall not discuss
this in the following.

It is convenient to consider the following functional \cite{5}
\beq{11}
M(r) = \frac{P^2_\Lambda}{2R} + \frac{R}{2} \left[ 1- 
       \left(\frac{R'}{\Lambda} \right)^2 \right].
\eeq
Making use of the constraints, it is straightforward to show that
\beq{12}
\frac{d M(r)}{dr} = - \frac{R'q^2}{2R^2}
\eeq
and thus
\beq{13}
M(r) = m - \frac{q^2}{2R(r)}.
\eeq
It is evident that this is just the total energy, with $m$ being the
ADM mass and $-q^2/2R$ the electrostatic energy, see e.g. \cite{9}.

We now assume the total state to be of the form (\ref{9}) and solve 
(\ref{10}) in a WKB approximation. Writing as usual 
$\psi \approx C e^{i S_0/\hbar}$ with a slowly varying prefactor $C$,
one finds for $S_0$ the Hamilton-Jacobi equation
\begin{equation} \label{14a}
\frac{G\Lambda}{2R^2} \left( \frac{\delta S_0}{\delta \Lambda} 
\right)^2
- \frac{G}{R} \frac{\delta S_0}{\delta \Lambda} 
\frac{\delta S_0}{\delta
R} + \frac{V_G}{G} + \frac{\Lambda q^2}{2 R^2} = 0 
\eeq
and the momentum constraint equation
\begin{equation}\label{14b}
R' \frac{\delta S_0}{\delta R} - \Lambda \frac{d}{dr} 
\frac{\delta S_0}{\delta \Lambda}= 0.
\eeq
In generalisation of \cite{3,4} one finds the solutions (up to a 
constant)
\begin{eqnarray}\label{15}
S_0 & = & \pm G^{-1}\int_{-\infty}^{\infty} dr \left\{ \Lambda Q -RR' 
\arcosh\frac{R'}{\Lambda \sqrt{1-\frac{2M}{R}}} \right\} \nonumber \\
& = & \pm G^{-1}\int_{-\infty}^{\infty} dr \left\{ \Lambda Q 
-\frac{1}{2}
  RR' \ln \frac{\frac{R'}{\Lambda} + \frac{Q}{R}}{\frac{R'}{\Lambda} 
  - \frac{Q}{R} } \right\}, 
\end{eqnarray}
where $Q$ is the functional
\beq{16}
Q \equiv R \sqrt{\frac{R'^2}{\Lambda^2} + \frac{2M}{R} -1 }.
\eeq
An analogous solution is found for two-dimensional dilaton gravity 
\cite{10} (see also \cite{11}).
Later we shall only consider the solution with the plus sign
in front of the integral. One may of course consider also 
superpositions of WKB states, but they will decohere after a 
coupling to other quantum fields is taken into account \cite{11}.
We note that the classical momenta found {}from (\ref{15}) read
\beq{17}
P_\Lambda \equiv \frac{\delta S_0}{\delta \Lambda} = \pm Q,
\hspace{1cm}
P_R \equiv \frac{\delta S_0}{\delta R} 
     = \pm Q^{-1} \left[\Lambda(m-R) +R\left( \frac{RR'}{\Lambda} 
       \right)'\right].
\eeq
Since $S_0$ can be considered as the generator of a canonical 
transformation, it is clear that spherically symmetric gravity can be 
classically reduced to a finite-dimensional system, since instead of 
arbitrary functions only the parameters $m$ and $q$ are contained
in (\ref{15}). This reduction has been explicitly done in \cite{5,6}.
The variables conjugate to $m$ and $q$ are obtained in the usual way 
{}from (\ref{15}) according to
\beq{18}
p_m = \frac{\partial S_0}{\partial m} = \mp \int dr 
      \left( 1-\frac{2M}{R}\right)^{-1} \frac{\Lambda Q}{R} ,
\eeq
\beq{19}
p_q = \frac{\partial }{\partial q}\left(S_0 + q\int dr \Gamma(r) 
      \right)
    = \pm \int dr \left\{ \left( 1- \frac{2M}{R} \right)^{-1} 
      \frac{\Lambda q Q}{R^2} + \Gamma \right\}.
\eeq
While $p_m$ describes the difference of parametrisation times at the 
left and right infinities of the Kruskal diagram \cite{5,6}, $p_q$
is related to the electromagnetic gauge choice 
\cite{7}.\footnote{Compare 
our expressions with the Eqs. (157) and (159)
in \cite{5} by use of $P_M=F^{-1}R^{-1}\Lambda P_\Lambda$ and the
Eqs. (4.3a) and (4.3b) in \cite{7}.}
As can be easily seen, $\psi\equiv \exp (i S_0/\hbar)$  
is an exact solution for the Wheeler-DeWitt equation
\beq{19a}
\left\{ -\frac{G\hbar^2 \Lambda}{2R^2}Q \frac{\delta }{\delta 
      \Lambda} 
      Q^{-1} \frac{\delta }{\delta \Lambda } + \frac{G\hbar^2}{R}Q 
      \frac{\delta }{\delta R} Q^{-1} \frac{\delta }{\delta \Lambda}
      +\frac{V_G}{G} + \frac{\Lambda q^2}{2 R^2} \right\} \psi =0,
\eeq
where a particular factor ordering has been chosen (compare also 
\cite{10} for a similar remark in the context of two-dimensional 
dilaton gravity). Wheeler-DeWitt quantisation of 
the constraints (\ref{2}) and (\ref{3}) by use of this particular 
factor ordering is thus equivalent to the quantisation of the reduced 
model. However, if another factor ordering in the Wheeler-DeWitt 
equation is chosen, going beyond the first WKB level would amount to
take into account a much wider class of three-geometries 
for consideration in the wave functional.
In this case the Wheeler-DeWitt approach can thus not be assumed to
be equivalent to the reduced approach to quantisation. Explicit 
calculations for higher order WKB terms have as yet been performed 
in most cases only in a formal sense \cite{12}.

We want to comment now on some interpretational issues with regard 
to (\ref{15}). Since classically $Q=P_\Lambda$ is real, $Q^2$ must be
positive, and therefore
\beq{20}
\left( \frac{R'}{\Lambda} \right)^2 \geq 1- \frac{2M}{R} 
\eeq
is the condition for the `classically allowed region'. We note that
for the classical spacetime metric, the condition $1-2M/R < 0$ 
describes the interior of the event horizon. One thus recognises 
from (\ref{20}) that this interior region is {\it always 
classically allowed}. We also note that in this region the logarithm
in (\ref{15}) acquires a term $i\pi$, since
\[\left( \frac{R'}{\Lambda} + \frac{Q}{R} \right) 
\left(\frac{R'}{\Lambda} - \frac{Q}{R} \right)= 1 - \frac{2M}{R} 
< 0.
\]
(The first factor becomes zero upon crossing the leftgoing horizon 
of the Kruskal diagram, while the second factor becomes zero upon 
crossing the rightgoing horizon \cite{5}.) In which sense this 
imaginary part is related to entropy will be discussed below. 
What are the classically forbidden regions? It is obvious that they
must correspond to three-geometries which {\it cannot} be embedded 
in a classical spacetime described by this model. Such 
three-geometries can, however, be embedded in the euclideanised 
classical spacetime. Since 
then $(R'/ \Lambda)^2 \le 1-2M/R$, the euclidean spacetime can cover
only regions outside the horizon, which is well known {}from the 
standard treatment \cite{1}.
As $\arcosh z = \pm i \arccos z$, the solutions (\ref{15}) for the 
classically forbidden region read (note that the argument of 
$\arccos$ is smaller than one)
\begin{eqnarray}\label{21}
S_0 &=& \pm i G^{-1} \int_{-\infty}^{\infty} dr \left\{ \Lambda R 
\sqrt{1-\frac{2M}{R}-\frac{R'^2}{\Lambda^2}} - RR' \arccos 
\frac{R'}{\Lambda\sqrt{1-\frac{2M}{R}}} \right\} \nonumber \\
    &=& \pm i G^{-1} \int_{-\infty}^{\infty} dr \left\{ \Lambda R 
\sqrt{1-\frac{2M}{R}-\frac{R'^2}{\Lambda^2}} - RR' \arctan 
      \frac{Q\Lambda}{RR'} \right\}, 
\end{eqnarray}
the latter being in agreement with the form of the generator with 
respect to the reduced euclidean model, see Eq.\ (6.10) in \cite{13}.

Since black hole entropy is recovered in the path integral 
formulation from a saddle point approximation \cite{1,2}, one 
should obtain it in the present framework from the 
Hamilton-Jacobi functional. In fact, it has been claimed that the 
entropy is related to the imaginary part of $S_0$ coming from the 
interior of the horizon in (\ref{15}) \cite{14}.
Inspecting first a three-geometry which, 
when embedded in the classical Kruskal spacetime, crosses both 
horizons, one recognises that in (\ref{15}) the contributions 
{}from the two horizon crossings to $\Im S_0$ cancel each other. 
There would thus be no candidate for the entropy. 
Such a cancellation was noted in \cite{11} for dilaton gravity, and 
is consistent with a similiar observation made in \cite{15} 
within the path integral framework.

Standard discussions of black hole thermodynamics often employ
three-geo\-me\-tries which orginate at the bifurcation two-sphere in
the Kruskal diagram \cite{13} or, more generally, the bifurcation 
surface of a Killing horizon \cite{16}. Is there a contribution to
the imaginary part of $S_0$ from such a boundary if the lower 
integration limit in (\ref{15}) is chosen to be this
region? The answer is {\it no}, since the three-geometry only covers
the region outside the horizon where no $i\pi$-term emerges from
the logarithm in (\ref{15}).

The crucial point in the recovery of the entropy is, however, the 
fact that boundary conditions at the bifurcation point lead to 
additional degrees of freedom \cite{17b,20}. 
This is fully analogous to the asymptotically flat case where 
additional degrees of freedom (there the generators of the 
Poincar\'e group) are present at spatial infinity \cite{17}. 
We assume in the following that the upper integration in (\ref{15}) 
corresponds to this situation \cite{5}.  For simplicity, we shall 
concentrate on the situation at the bifurcation sphere (where we 
assume for $r$ the value $r=0$) and will make use of the equations 
found by Kucha\v{r} in \cite{5} for the upper 
boundary.\footnote{Generalisations to 
black holes embedded in a box \cite{13} or in an anti-de Sitter 
spacetime \cite{7} can easily be done.}

What are the boundary conditions at the bifurcation sphere?
They are chosen in such a way that the classical solutions have a 
nondegenerate horizon, and that the hypersurfaces $t=constant$ begin 
at $r=0$ in a manner asymptotic to hypersurfaces of constant Killing
time \cite{7,13}.
In particular, one has near $r=0$:
\beq{21b}
N(t,r) = N_1(t)r+ O(r^3),
\eeq
\beq{21c}
\Lambda(t,r) = \Lambda_0(t) + O(r^2),
\eeq
\beq{21d}
R(t,r)= R_0(t) + R_2(t) r^2 + O(r^4),
\eeq
where $R_0 \equiv R(0)$ is defined by $(1-2M/R)|_{r=0}=0$.
Note that in (\ref{21b}) $N_1$ is only non-vanishing if $N$ has a 
single root at the origin. In the case of a double root, 
$\partial N/\partial r =0 =N_1$. This will become important for the
extreme Reissner-Nordstr\"om case. Employing the above boundary 
conditions (and similar ones for the conjugate momenta and the 
shift function \cite{7,13}), one recognises that the variation of 
the classical action (in the following we shall neglect
the $\Gamma$-part)
\beq{22}
S_\Sigma[\Lambda,P_\Lambda,R,P_R;N,N^r] =\int dt \int_{0}^{\infty} dr 
(P_\Lambda \dot{\Lambda} + P_R \dot{R} - N {\cal H}_G - N^r 
{\cal H}_r)
\eeq
leads to the following term at $r=0$:
\beq{23}
\left. \delta S_\Sigma \right|_{r=0} 
= - \frac{\partial }{\partial r} \left. \left( N 
    \frac{\partial {\cal H}_G}{\partial R''}\right) \delta R  \right|_{r=0}
= - \frac{N_1R_0}{G \Lambda_0} \delta R_0.
\eeq
If $N_1 \not= 0$ one must subtract this boundary term from the
original action (\ref{22}). Otherwise the variation of the action
with respect to $R$ would lead to the unwanted conclusion that 
$N_1/\Lambda_0 = 0$.
This is analogous to the situation at infinity 
where one has to add the ADM energy term \cite{5,17}. One thus has 
to consider the classical action
\begin{eqnarray}\label{24}
S[\Lambda,P_\Lambda,R,P_R;N,N^r] &=&
   \int dt \int_{0}^{\infty} dr ( P_\Lambda \dot{\Lambda} + P_R 
   \dot{R} - N {\cal H}_G - N^r {\cal H}_r) \nonumber \\
 & & + \frac{1}{2} \int dt \frac{N_1 R_0^2}{G \Lambda_0}
     - \int dt N_+ M_+,
\end{eqnarray}
where $M_+$ denotes the ADM energy and $N_+$ the lapse function at
infinity.
The boundary term $\frac{1}{2G} \int dt  R_0^2  \delta\! 
\left(\frac{N_1}{\Lambda_0}
\right)$ vanishes if one assumes that $N_1/\Lambda_0\equiv N_0$ is 
fixed at $r=0$ \cite{13}. If $N_1=0$, no boundary term emerges and
one is left with the original action (\ref{22}). This happens in the
case of extreme Reissner-Nordstr\"om black holes. A similar 
conclusion has been reached from a euclidean viewpoint in 
\cite{20}.

The necessity of fixing $N_0$ at $r=0$ and $N_+$ at 
infinity can be removed by introducing the parametrisations
\beq{25}
N_0(t) = \dot{\tau}_0(t), \hspace{1cm} N_+(t)= \dot{\tau}_+(t)
\eeq
with $\tau_0$ and $\tau_+$ as additional variables. Instead of 
(\ref{24}) one then considers
\begin{eqnarray}\label{26}
S[\Lambda,P_\Lambda,R,P_R;N,N^r] &=&
   \int dt \int_{0}^{\infty} dr ( P_\Lambda \dot{\Lambda} + P_R 
   \dot{R} - N {\cal H}_G - N^r {\cal H}_r) \nonumber \\
 & & + \int dt \left(\frac{R_0^2}{2G} \dot{\tau }_0 - M_+ 
     \dot{\tau}_+ \right). 
\end{eqnarray}
{}From the canonical point of view, this does not yet yield a 
satisfactory description, since there are no momenta canonically 
conjugate to $\tau_0$ and $\tau_+$. 
One can interpret the action (\ref{26}) as representing a mixed 
Hamilton-Lagrangian form. In order to introduce the canonical momenta
$\pi_0$ and $\pi_+$, one has to perform a standard Legendre 
transformation.
But this can only be consistently done, if two new constraints are 
introduced: 
\beq{27} 
{\cal C}_0 \equiv \pi_0 -\frac{R_0^2}{2G} \approx 0,
\eeq
\beq{27b}
{\cal C}_+ \equiv  \pi_+  + M_+  \approx 0.
\eeq
These constraints must be adjoined to the action by Lagrange 
multipliers $N_+$ and $N_0$:
\beq{27c}
S[\Lambda,P_\Lambda,R,P_R;\tau_0,\pi_0,\tau_+,\pi_+;N,N^r,N_0,N_+] 
\eeq
\begin{eqnarray}
&=& \int dt \int_{0}^{\infty} dr ( P_\Lambda \dot{\Lambda} + P_R 
    \dot{R} - N {\cal H}_G - N^r {\cal H}_r) \nonumber \\
& & + \int dt \left( \pi_0 \dot{\tau}_0 + \pi_+ \dot{\tau}_+ - N_0 
    C_0 - N_+ C_+ \right). \nonumber
\end{eqnarray}
The new form (\ref{27c}) of the action gives rise to additional terms
in the WKB approximation arising from implementing the constraints 
(\ref{27}) and (\ref{27b}) on wave functions:
\beq{27d}
\frac{\partial S_0}{\partial \tau_0} - \frac{R_0^2}{2G} = 0,
\eeq
\beq{27e}
\frac{\partial S_0}{\partial \tau_+}  +  M_+  =0,
\eeq
which alter the above solution $S_0$ of the Hamilton-Jacobi equation 
by
\beq{29}
S_0 \longrightarrow S_0 + \frac{R_0^2 }{2G}\tau_0 - M_+ \tau_+.
\eeq
To recover from this expression a term which can be interpreted as 
an entropy requires an appropriate euclideanisation in order to make 
contact with the formalism involving partition functions. 
The standard transition to the
euclidean regime uses $N_0 \equiv -i N_0^E$ to obtain a well-defined 
partition function from the path integral, where $N_0^E$ is the 
euclidean lapse function in the line element
\beq{30}
ds_E^2= (N_0^E)^2 \tilde{r}^2 dt^2 + d\tilde{r}^2 + R^2 d\Omega^2
\eeq
with $d\tilde{r} = \Lambda dr$ (and $N^r$=0).
Thus, $ \tau_0 \rightarrow -i \int dt N_0^E \equiv -i \tau_0^E$, 
which means that one has to choose $\tau_0^E= 2\pi$ because the 
`time'-integration (\ref{30}) is around the circle $S^1$, and 
$\tau_0^E$ is the angle.
One thus arrives at the `euclidean' WKB-state
\beq{31}
\psi[\Lambda(r),R(r);\beta, \tau_0=2\pi) 
   = \exp \left\{-\frac{1}{\hbar} S_0^{E} - \frac{\beta}{\hbar}  M_+ 
     + \frac{\pi R_0^2}{\hbar G}  \right\},
\eeq
where $S_0^E$ is $(-i)$ times the expression (\ref{21}). 
One can interpret
$\beta^{-1}=-i{\cal T}^{-1}$ with ${\cal T} \equiv \int dt N_+(t)$ 
as the renormalised temperature at infinity as in \cite{7},
and $\pi R_0^2/\hbar G \equiv A/4\hbar G$ as the Bekenstein-Hawking
entropy. 
Since no boundary term at the bifurcation two-sphere arises in the
extreme Reissner-Nordstr\"om case, there is no entropy in that case.
Note that for the above derivation it was not necessary to
perform explicit lapse-redefinitions such as in \cite{13}.
The result (\ref{31}) is in full analogy to the result from the 
euclidean path integral, where the entropy arises from the surface 
term in the Einstein-Hilbert action \cite{1}.

The above derivation suggests the following viewpoint with regard to 
the recovery of the entropy from WKB quantum states: 
For three-geometries which in the classical spacetime correspond to 
slices through the full Kruskal diagram, the `information' is maximal
in the sense that data on such a slice allow one to recover the full 
spacetime. This point was also made in \cite{15} to explain the 
vanishing of the entropy, which was found from the path integral for
such slices. For slices  which start at the bifurcation sphere, the 
information is less than maximal for the Schwarzschild as well as 
for the non-extreme ($q^2<m^2$) Reissner-Nordstr\"om black holes.
Therefore they are attributed the entropy $A/4\hbar G$.
In the extreme Reissner-Nordstr\"om case ($q^2=m^2$) the maximum 
information is already available for such slices (compare the Penrose
diagram of this case \cite{18}).
Therefore its entropy is zero, in accordance with results obtained 
from the euclidean formalism \cite{20,19}.
We note that the discontinuity of entropy in the extremal limit may be
problematic in the framework of string theory \cite{0a}, since the solution 
with vanishing entropy may be unstable. 

An interesting result is obtained in the Schwarzschild case for a
three-geometry that starts at the singularity, crosses one of the 
horizons and goes to infinity.
In this case one cannot assume the euclidean viewpoint, since, as has
been discussed above, there is no interior region in this case. 
However, $S_0$ in Eq.\ (\ref{15}) now acquires an imaginary part -- 
leading to a real part in the exponential of the WKB state -- from 
the interior region where the imaginary term of the logarithm is 
$i\pi$. This yields
\beq{32}
\Im S_0 =-\frac{1}{2G} \int dr RR' \pi = - \frac{\pi}{2G} 
          \int_0^{R_0} dR \; \frac{1}{2} R^2 = -\frac{\pi R_0^2}{4G}.
\eeq
Since $\psi \approx \exp( i \Re S_0 -\Im S_0)$, this corresponds
to an entropy of $\pi R_0^2/4G$, one fourth of the Bekenstein-Hawking
entropy. This lower value arises because now part of the interior 
regions can also be recovered from initial data on such slices in 
the classical theory, corresponding to `more information'.

In summary, we have shown that black hole entropy can be recovered 
from WKB quantum states in a natural way. We were able to deduce 
the Bekenstein-Hawking entropy by taking into account additional 
degrees of freedom.
These additional degrees of freedom arise by inspecting
the boundary conditions for particular three-geometries after they
are interpreted as embeddings in the underlying Kruskal diagram of
the classical theory.
The connection between these degrees of freedom and the entropy of 
black holes was made through a euclideanisation of the classical 
line element. 
This seems to suggest that the Bekenstein-Hawking entropy can only 
be derived from WKB quantum states after a Wick rotation has been
accomplished.
For this reason we are not able to obtain a discrete mass spectrum 
of the black hole by using the periodicity of the Wick-rotated
coordinate, in contrast to \cite{21}. 
Note that in our approach this periodicity is needed to interpret
part of the Hamilton-Jacobi functional as the entropy of the black
hole in the first place.
On the other hand, we could follow a recent suggestion \cite{21a} and
use the periodicity of $\tau_0$ to demand the validity of 
Bohr-Sommerfeld conditions
for the canonical pair $(\tau_0, \pi_0)$ in the euclideanised 
spacetime,
\beq{33}
n h = \oint \pi_0 d\tau_0 = \int_0^{2\pi} \frac{R_0^2}{2G} d\tau_0
    = \frac{4\pi M^2}{G}.
\eeq
This leads to the same  discrete spectrum for the mass as in 
\cite{22}. 
Whether such a quantisation of the black hole entropy also holds
in the physically relevant lorentzian case is an open 
issue and subject to future investigations.

\acknowledgments

We are grateful to Heinz-Dieter Conradi and Miguel Ortiz for
critical comments.
One of us (T.B.$\!$) acknowledges financial support from the 
{\it DFG-Graduierten\-kol\-leg Nichtlineare Differentialgleichungen
der Albert-Ludwigs-Universit\"at Freiburg}. 


\end{document}